\documentclass{article}

\usepackage[utf8]{inputenc} 
\usepackage[T1]{fontenc}    
\usepackage{hyperref}       
\usepackage{url}            
\usepackage{booktabs}       
\usepackage{amsfonts}       
\usepackage{nicefrac}       
\usepackage{microtype}      
\usepackage{xcolor}         

\usepackage{amsmath}
\usepackage{amssymb}
\usepackage{amsthm}
\usepackage{mathtools}

\usepackage{graphicx}
\usepackage{psfrag}
\usepackage{authblk}
\usepackage{caption}
\usepackage{subcaption}
\usepackage{url}

\usepackage{xtab,booktabs}

\newcommand{\R}{\mathbb R}

\numberwithin{theo}{section}

\usepackage[preprint,nonatbib]{mlncp_2023}

\author{%
Thomas Geert de Jong$^{1}$ \;\; Nozomi Akashi$^{2}$ \;\; Tomohiro Taniguchi$^3$ \vspace{-3mm} \\ 
\;\; Hirofumi Notsu$^1$ \;\; Kohei Nakajima$^4$ \vspace{1mm} \\ 
$^1$Kanazawa University \quad $^2$Kyoto University \\ $^3$National Institute of Advanced Industrial Science and Technology \quad $^4$University of Tokyo \vspace{1mm}\\
\texttt{\{tgdejong,notsu\}@se.kanazawa-u.ac.jp}, \texttt{akashi.nozomi.2a@kyoto-u.ac.jp,} \\ \texttt{tomohiro-taniguchi@aist.go.jp,} \texttt{k-nakajima@isi.imi.i.u-tokyo.ac.jp}
}

\title{Virtual reservoir acceleration for CPU and GPU: Case study for coupled spin-torque oscillator reservoir
}
\begin{document}
\maketitle

\begin{abstract}
 We provide high-speed implementations for simulating reservoirs described by $N$-coupled spin-torque oscillators. Here $N$ also corresponds to the number of reservoir nodes. We benchmark a variety of implementations based on CPU and GPU. Our new methods are at least 2.6 times quicker than the baseline for $N$ in range $1$ to $10^4$. More specifically,  over all implementations the best factor is 78.9 for $N=1$ which  decreases to 2.6 for $N=10^3$ and finally increases to 23.8 for $N=10^4$.  GPU outperforms CPU significantly at $N=2500$. Our results show that GPU implementations should be tested for reservoir simulations. The implementations considered here can be used for any reservoir with evolution that can be approximated using an explicit method. 

\end{abstract}


\section{Introduction}

With the recent advancements in machine learning, such as deep learning and generative models, the computational demands and associated energy costs have rapidly escalated~\cite{emma2019energy, thompson2020computational}.
Consequently, there is an increasing emphasis on the development of more efficient computers to perform machine learning tasks.
Computers that emulate neural networks in hardware are referred to as neuromorphic devices and have garnered attention as next-generation efficient computers.
Physical reservoir computing is one of the major frameworks implemented in neuromorphic devices, which exploits the dynamics of optic, quantum, and spintronics systems \cite{nakajima20}. Their evolution is represented as a coupled network with a massive amount of nonlinear nodes called the \textit{reservoir}. Only the readout part has to be trained for the learning task which makes it extremely computationally efficient. However, finding optimal physical parameters or number of nodes for the reservoir can be a time-consuming effort. In addition, the information processing capabilities of the reservoir increase generically with the size \cite{dambre12,kubota21}. Hence, to aid in this endeavour simulations of the reservoir are performed. But even in a virtual environment where we consider large number of nodes over a parameter space it becomes again a time-consuming effort. Although high speed implementations for toy reservoirs such as ESN exist \cite{echotorch}, we think that speeding up current simulators of physical reservoirs described by ODEs and PDEs would significantly contribute to the research community.

This study is done by example. Here we consider coupled spintronic devices. Spintronic devices exhibit excellent characteristics, making them promising candidates for neuromorphic computers, including high-speed dynamics, minute size, high energy efficiency, and durability against radiations~\cite{patten1996overview, gerardin2010present, Kudo_2017, romera2018vowel, grollier20}.
A type of spintronic device known as a Spin-Torque Oscillator (STO) shows diverse dynamics, including periodic oscillation, and transitions between fixed points and chaos~\cite{yang07, williame19, taniguchi19, yamaguchi19, akashi20, kamimaki21}.
It has been reported that these dynamics can be directly harnessed as computational resources using physical reservoir computing schemes~\cite{torrejon17, furuta2018macromagnetic, tsunegi18, tsunegi19, jiang19, kanao19, markovic2019reservoir, yamaguchi20, akashi20, akashi22, tsunegi2023information, nakajima20}.

\begin{figure}[ht]
\centering
\includegraphics[width=10cm]{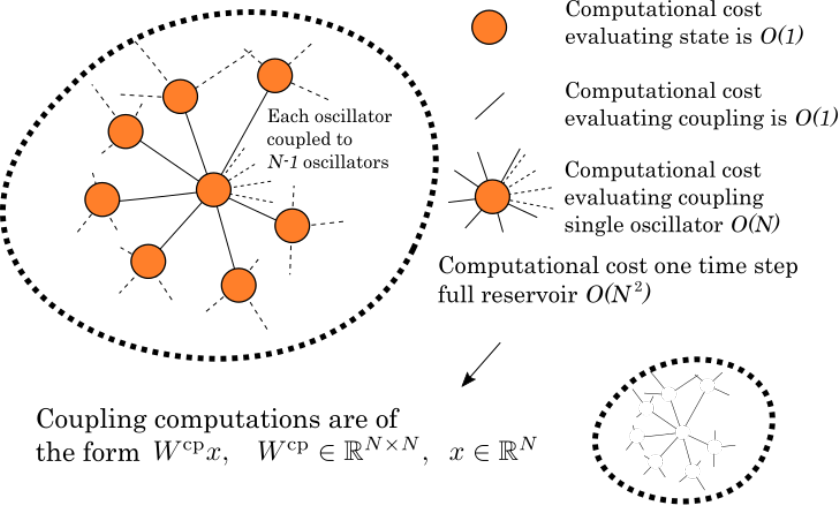}
\caption{$N$-coupled STO simulations are suitable for parallelization for sufficiently large $N$ as the coupling computations are matrix multiplications.}
\end{figure}

In this paper we benchmark CPU and GPU  implementations for simulating reservoirs given by $N$-coupled STOs. The GPU implementation is motivated by the fact that the coupling terms are of complexity $O(N^2)$ which can run quickly on GPUs due to their high performing parallelization ability. Each implementation is optimal in some $N$ range.  We show that the $N$ used in the literature is sufficiently large that GPUs are quicker than CPUs. The methods considered here are general and can be applied to any reservoir which can be approximated using an explicit method.  This paper is accompanied by a github repository with the benchmarking code \cite{reservoir_acceleration}.

\section{Background \label{sec:lit}}


\textbf{GPU-deployed reservoirs.} In \cite{gallicchio2017deep,echotorch} a GPU-deployment of Echo State Network (ESN) is studied with a focus on stacking ESNs. However, no ablation studies are given of CPU versus GPU computations. In addition, ESNs are not described by differential equations and ESN do not have a physical representation.    

\textbf{ODE solving with GPUs.} Neural networks have made impressive contribution to a new range of ODE solvers \cite{chen2018neuralode, chen2018neural, greydanus2019hamiltonian,cranmer2020lagrangian}. Although their performance is incredible, these results typically lack rigorous statements concerning convergence and error estimates. As a consequence neural net approaches are not widely adopted by standard solvers, for example COMSOL, FreeFEM, Matlab, Mathematica, still rely on conventional methods that are deployed on CPUs.  \cite{torchdiffeq} contains some explicit solvers for GPU which would be more efficient than CPU if the system is sufficiently large and parallizable. In \cite{magnum23} a PyTorch based micromagnetic simulation library is presented. The repository contains code for simulating a single spin-torque oscillator. There is no direct way to introduce the coupling between the oscillators without tearing the code apart. 

\textbf{Reservoir computer parameters and number of nodes.} In theoretical reservoir studies which focus on predicting complicated chaotic time series the coupled components are typically of order $O(10^3)$ \cite{jaeger04,gauthier2021next, pathak2018model}. For a physical reservoir this is not within the realm of possibilities and one typically only tests 1 device \cite{tsunegi18,torrejon17}. Generally, in a virtual setting improving reservoir computer performance and understanding the underlying dynamics requires an exploration of the parameter space \cite{lu2018attractor,flynn2021multifunctionality,akashi20,pathak2017using}. As the reservoir computer has to be trained and tested for each parameter this is a computationally expensive task.

\section{Methodology}

\subsection{Coupled STO equations}

We consider $N$-coupled STOs. The differential equations corresponding to the magnetization of the $k$th-STO, $\mathbf{m}_k(t) = (m_k^x(t), m_k^y(t), m_k^z(t))^\top$, can be obtained through the Landau--Lifshitz--Gilbert~(LLG) equations. In Cartesian co-ordinates we can express them as the following first order ODE: 
\begin{align}
\frac{d \mathbf{m}_k}{dt} =  - \frac{\gamma}{1+\alpha^2}  \mathbf{m}_k \times \mathbf{b}_k(\mathbf{m},\textbf{u}) - \frac{\alpha \gamma}{1+ \alpha^2} \mathbf{m}_k \times (\mathbf{m}_k \times \mathbf{b}_k(\mathbf{m},\textbf{u}) ) \label{eq:gov}
\end{align}
where $\mathbf{b}_k(\mathbf{m},\textbf{u})= \mathbf{H}^{\rm total}_k(\mathbf{m},\textbf{u}) + H_{s}(\mathbf{m}_{k}) \mathbf{p} \times \mathbf{m}_k$ with ${H}_{s}(\mathbf{m}_{k}) =  \frac{\hbar \eta I }{2e(1+ \lambda \mathbf{m}_k \cdot \mathbf{p} )MV}$ and the total field $\mathbf{H}^{\rm total}_k$ given by the sum of the effective magnetic field, spin-transfer torque, coupling field, and input field, $\mathbf{H}^{\rm total}_k(\mathbf{m},\mathbf{u}) =  \mathbf{H}(\mathbf{m}_k)  + \mathbf{H}^{\rm cp}_k (\mathbf{m}) + \mathbf{H}^{{\rm in}}_k(\mathbf{u}) $ with    
\begin{align}
 \mathbf{H}(\mathbf{m}_k) &= [H_{\rm appl} + (H_K - 4 \pi M)m_{k}^z]\mathbf{e}_z ,  \nonumber \\
 \mathbf{H}^{{\rm cp}}_k (\mathbf{m}) &= A_{\rm cp} \sum_{i=1}^N w^{\rm cp}_{k,i} m_{k}^x\mathbf{e}_x, \label{eq:coup} \\
 \mathbf{H}_k^{\rm in}(\mathbf{u}) &= A_{\rm in} \sum_{i=1}^{N_{\rm in }}w^{\rm in}_{k,i} u_i \mathbf{e}_x. \label{eq:in}
\end{align}
The input signal $\textbf{u}(t)=(u_1(t), u_2(t), \ldots , u_{N_{\rm in}}(t))^\top$ is a discrete-point series. Let $W^{\rm cp} \in {\R}^{N \times N}$ denote the coupling matrix given by $(W^{\rm cp})_{k,i} =  w^{\rm cp}_{k,i}$. There is no self-coupling, $(W^{\rm cp})_{k,k} =0$, and $(W^{\rm cp})_{k,i}$ with $k\neq i$ sampled from a uniform distribution on $[-1,1]$.  The spectral radius of $W^{\rm cp}$ is set to 1. The components $w^{\rm in}_{k,i}$ are sampled from a uniform distribution on $[-1,1]$. A summary of parameters is given in Table \ref{tab:sum_not}.  The parameters are chosen such that the solution exhibits oscillatory motion.  These parameter values have been used in various reservoir experiments (virtual and physical) \cite{kubota2013spin,taniguchi2017relaxation}.

\begingroup
 \begin{table}
\centering
 \begin{tabular}{lll}
\toprule
Symbol & Definition & Value \\
\midrule
$\gamma$ & gyromagnetic ratio & $1.764 \cdot 10^7$ rad/Oe s \\
$\alpha$ & Gilbert damping constant & 0.005\\
$M$ & saturation magnetization & 1448.3 emu/cm$^3$\\
$H_{\rm K}$ & interfacial magnetic anisotropy field & 18.616 kOe \\
$H_{\rm appl}$ & applied field & 200 Oe\\
$\eta$ & spin polarization & 0.537\\
$\lambda$ & spin-transfer torque asymmetry & 0.288\\
$I$ & electric current & 2.5mA\\
$V$ & volume of free layer & $\pi \cdot 60^2 \cdot 2$ nm$^3$\\
$e$ & elementary charge & $1.60217733 \cdot 10^{-19}$ C \\
$\hbar$ & Planck constant, reduced &  $1.05457266 \cdot 10^{-34}$ Js\\
$\mathbf{p}$ &  perturbed direction of pinned layer magnetization &  $(1, 0,  6.123234 \cdot 10^{-17})^\top$\\ 
$ A_{\rm cp}$  & coupling amplitude &  1 Oe \\
$A_{\rm in}$ & input amplitude &  1 Oe \\
 \bottomrule
\end{tabular}
\caption{Summary of parameters\label{tab:sum_not}}
\end{table}
\endgroup
We note that the state space is 3$N$-dimensional. Typically, $N$-states are used as the nodes of the reservoir. For initial conditions we consider 
\begin{align}
\mathbf{m}_k(0) = (\sin(\phi_0)\cos(\phi_0),\sin(\phi_0)\sin(\phi_0),\cos(\phi_0))^\top,    \label{eq:initial}
\end{align}
with $\phi_0=2 \pi/360$. Observe that $\mathbf{m}_k(0)\approx (0,0,1)^\top$ and that $|\mathbf{m}_k(0)|=1$. Straightforward calculation shows that $|\mathbf{m}_k|$ is conserved~\cite{bertotti09}. Then, from \eqref{eq:initial} it follows that
\begin{align}
|\mathbf{m}_k| = 1.
\end{align}

\begin{figure}[ht]
\centering
	\begin{subfigure}[b]{0.45\textwidth}
	\centering
	\includegraphics[width = 5cm]{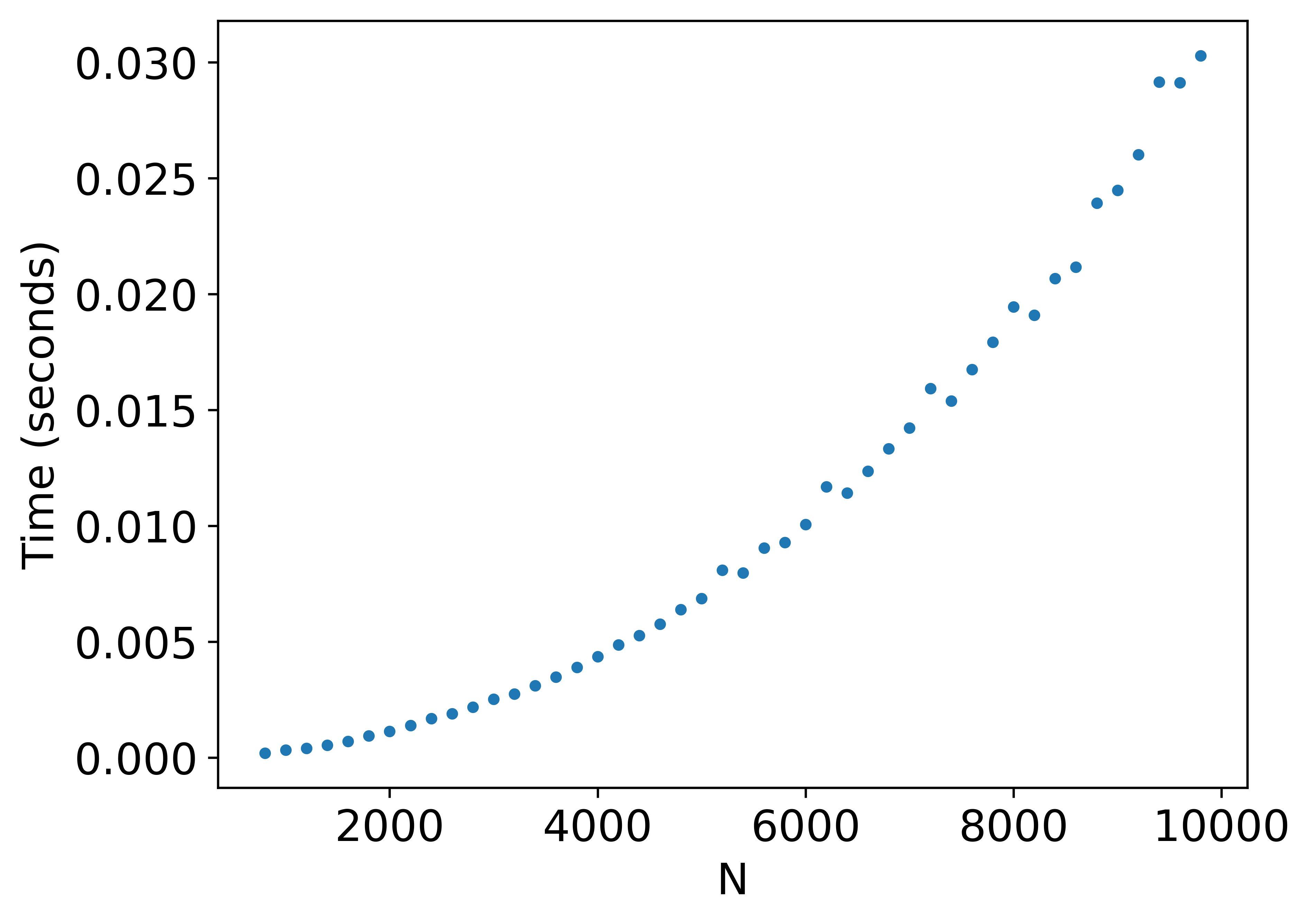}
        \caption{Linear scale \label{fig:lin}}
	\end{subfigure}
    \begin{subfigure}[b]{0.45\textwidth}
	\centering    
    \includegraphics[width= 5cm]{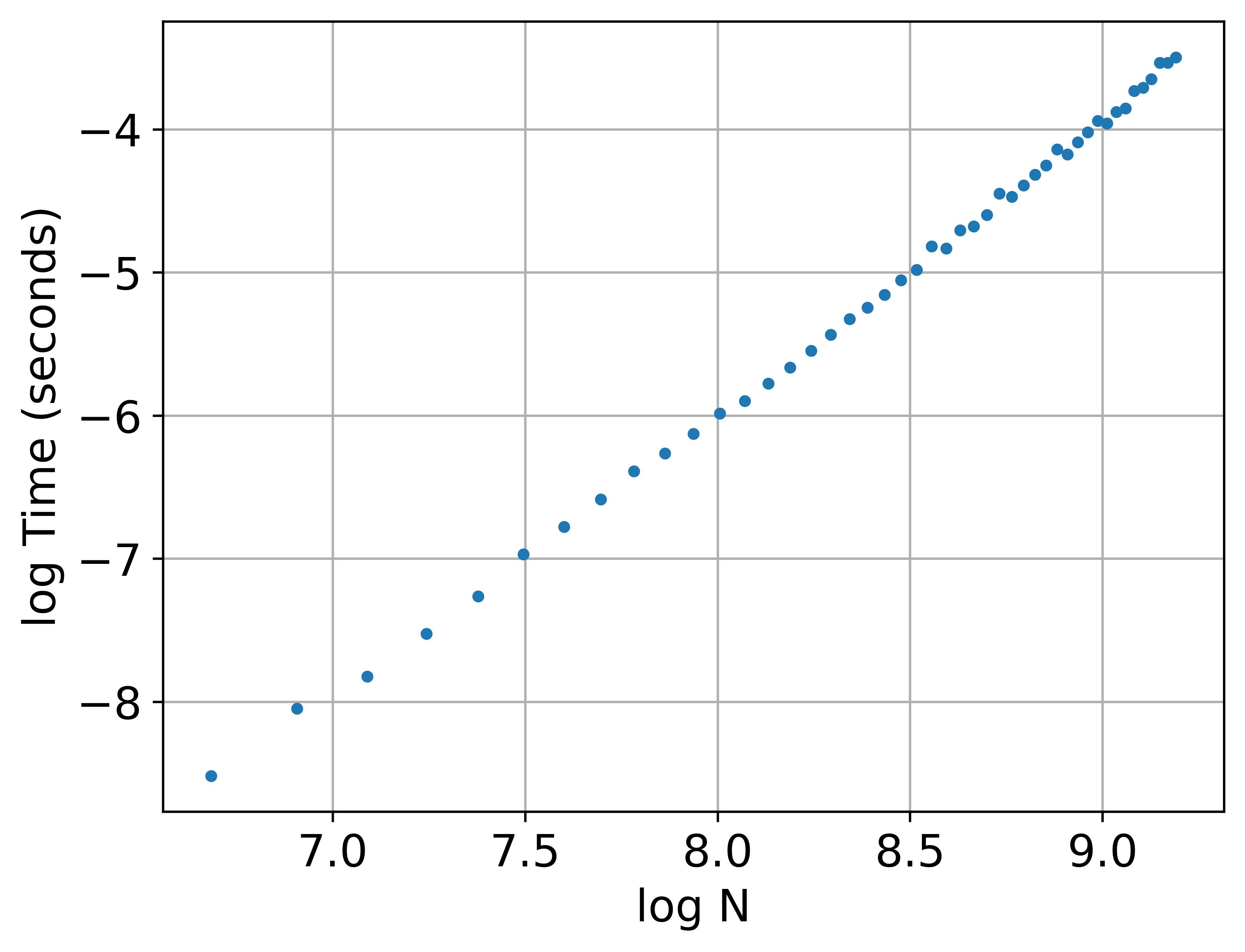}
    \caption{Log scale \label{fig:log}}
    \end{subfigure}
    \caption{Computation time vector field for random initialization of $\mathbf{m}$  \label{fig:compu_vec}}
\end{figure}

Straightforward computations shows that evaluation of the vector field \eqref{eq:gov} is $O(N^2)$. Numerically we can validate this by showing that the computation time of the vector field for random initialization increases quadratically in $N$, see Figure \ref{fig:compu_vec}

\subsection{Benchmark test}

For the benchmark test we set the input, $\mathbf{u}$, equal to zero as the input field term~\eqref{eq:in} is $O(1)$. We observe that the coupling field term~\eqref{eq:coup} is $O(N)$. Observe that if the coupling field is set to 0 the vector field can be evaluated in $O(N)$. Hence, for meaningful results the coupling field has to be incorporated in the benchmark test. We fix the precision to float64.

Based on Section~\ref{sec:lit} we consider the $N$-range 1 to 10 000.

We solve Equation~\eqref{eq:gov} using (classic) 4th order Runge--Kutta with step size $10^{-11}$ for $5 \cdot 10^{5}$ steps. To validate the correctness for a given $N$ we check for each implementation whether the solutions and the conservation law $|\mathbf{m}_i| = 1$ are identical.

\subsection{CPU and GPU implementations}

We consider the following Python implementations for the benchmark test: 
\begin{itemize}
\item[-] \textbf{CPU, Numpy (Base)}: vectorized Numpy, 
\item[-] \textbf{CPU, Numba-vanilla}: straightforward Numba~\cite{lam2015numba},  
\item[-] \textbf{CPU, Numba-parallel}: parallelization with Numba,
\item[-] \textbf{GPU, Torch}: Pytorch tensor running on CUDA.
\end{itemize}
We provide additional details to the methods. Here we use community provided Numpy code as a baseline. Numba-vanilla is a fully sequential Numba implementation. When running Numba-vanilla we observed that the coupling field~\eqref{eq:coup} is a bottle neck as $N$ increases. Hence, in Numba-parallel we optimize the coupling field computation by parallelization. Over the full vector field the coupling field can be written as a matrix calculation of the form $W^{cp}x$ with $x \in \R^N$. In Numba-parallel only the coupling field computation is parallelized.  Finally,  Numba acts as a JIT (Just-In-Time)  compiler. We chose to compile the code prior to benchmarking. 

We finally note that the order of performing operations differs for each implementation. This leads to a machine precision difference in the approximation of the solutions at the first time step. The differences between methods will grow as the number of steps are increased. However, this difference is at least a factor $10^{-6}$ smaller than the error corresponding to the preserved quantity $|\mathbf{m}_i|-1$. The repository \cite{reservoir_acceleration} contains supplemental experiments.

\subsection{Hardware and module versions}
The CPU and GPU used for these experiments are AMD Ryzen 9 5950X (16-core) and Nvidia RTX A4500, respectively. 

We use Numpy v1.24, Numba v0.57, Torch v2.0. We note that Numpy v1.24 is the most recent Numba-compatible version.

\section{Results}

In Table \ref{tab:seconds} we present the computation times for the benchmark test for different $N$ and in Table \ref{tab:factor} we present the factor improvement over the Numpy base which is an improvement over implementations used in the literature~ \cite{akashi22}. We note that Table~\ref{tab:seconds} concerns averaged results. However, deviations over multiple runs are small. For all computations the standard deviation is at least less than 2 orders of the leading decimal of the mean.

For small $N$ Numba-vanilla has the quickest computation times with a factor improvement of $O(10)$ over the base. For $N$ sufficiently large Numba-parallel performs better than the base with best computation times at $N=1000$ with respect to the other implementations. Finally, the performance boost through parallelization is most visible for the GPU which beats the CPU for $N\geq 2500$. The GPU hits a factor $O(10)$ improvement over the base as $N$ is increased. 

\begingroup
 \begin{table}
\centering
 \begin{tabular}{lcrrrrrrr}
\toprule
 &  &   \multicolumn{7}{c}{$N$}  \\
 \cmidrule(lr){3-9}
CPU/GPU &  Method & 1  &  10 & 100  & 1000 & 2500 & 5000 & 10000 \\
\midrule
CPU &  Numpy (Base) &  0.81s& 0.93s& 1.12s& 5.78s& 36.38s & 2.79m &  13.40m\\
\midrule
CPU & Numba-vanilla & \textbf{0.010}s& \textbf{0.016}s& \textbf{0.14}s& 12.56s& 86.23s& 5.99m & 23.66m     \\
 CPU & Numba-parallel &X& 0.48s& 0.60s&  \textbf{2.27}s& 21.59s& 2.60m &  11.54m\\
 \midrule 
GPU & Torch & X & 7.47s & 7.55s &  7.14s & \textbf{7.59}s & \textbf{0.22}m &  \textbf{0.56}m\\ 
 \bottomrule
\end{tabular}
\caption{Computation time. Here s denotes seconds and m denotes minutes. For each $N$ the best computation time is denoted in bold. \label{tab:seconds}}
\end{table}
\endgroup

Numba-vanilla does not have any in-built method to deal with large systems, contrary to Numpy. Hence, it only performs better than the Numpy base for small systems. For $N\geq 10^3$ the overhead of parallelization becomes cost-effective as Numba-parallel beats the other implementations. This argument also applies to the GPU results. For small matrix computations the GPU cannot function optimally which might be the cause for the non-increasing computation times occurring for $N \leq 10^3$.

\begingroup
\begin{table}
\centering
\begin{tabular}{lcrrrrrrr}
\toprule
 &  &   \multicolumn{7}{c}{$N$}  \\
 \cmidrule(lr){3-9}
CPU/GPU &  Method & 1  &  10 & 100   & 1000 & 2500 & 5000 & 10000 \\
\midrule
CPU & Numba-vanilla & \textbf{78.9}  &  \textbf{57.9}  &\textbf{ 7.9}    & 0.5  & 0.4   & 0.5  &  0.6\\
CPU & Numba-parallel &X&  1.9 & 1.9 &  \textbf{2.6}  & 1.7 & 1.1  & 1.2  \\
 \midrule 
GPU & Torch & X & 0.1 & 0.1  &   0.8 & \textbf{4.8} &  \textbf{12.9}  & \textbf{23.8}  \\ 
 \bottomrule
\end{tabular}
\caption{Speed-factor with respect to base (Factor = time base / time method). For each $N$ the best factor is denoted in bold.  \label{tab:factor}}
\end{table}
\endgroup


\section{Conclusion and future work}

In this work we present implementations for speeding up coupled STO simulations for reservoir computing. We obtain a speed factor improvement of at least 2.6 which goes up to $O(10)$ for $N \sim O(10)$ and $O(10^4)$. However, the implementations for CPU and GPU considered in this paper apply to any reservoir whose evolution can be approximated using an explicit method. This paper is accompanied by a github respository containing the benchmarking code.

An important result is that GPU based reservoir computing should be considered when improving computation times. Overhead for GPU computations is high.  Hence, \textit{a priori} it is unclear if GPU usage is meaningful for differential equations. Our work shows by example that in a virtual setting the reservoir nodes are of an order that GPU implementations should be tested for performance. 

By removing reservoir nodes and artificially replacing them using a delay-operation, such as multiplexing, the computational time can be reduced.  However, this does not necessarily increase the information processing capabilities of the reservoir. Hence, it is preferable to develop numerical methods that speed up the simulation of natural reservoirs. This work contributes to this direction.

The physical implementation of reservoirs with a large number of interacting components is likely not a topic that can be studied in the near future. However, by considering a spatial and discretized system, i.e. moving from ODEs to PDEs, we can efficiently construct high dimensional reservoirs. A fluid is a suitable candidate for this approach. Pioneering work was accomplished in \cite{goto2021twin} by achieving the first vortex (virtual) reservoir computer. The underlying CPU-based computations are computationally intensive (order of days) and would greatly benefit from a GPU-based approach.

\acksection{This research was supported by JSPS KAKENHI Grant Number 22J01542 and JST CREST Grant Number JPMJCR2014.}

\bibliographystyle{alpha}
\bibliography{abbriv.bib}{}

\end{document}